\documentclass[conference]{IEEEtran}
\IEEEoverridecommandlockouts
\usepackage{cite}
\usepackage{amsmath,amssymb,amsfonts}
\usepackage{algorithmic}
\usepackage{graphicx}
\usepackage{textcomp}
\usepackage{xcolor}
\usepackage{booktabs}
\usepackage{comment}
\usepackage{enumerate}
\usepackage{threeparttable}
\def\BibTeX{{\rm B\kern-.05em{\sc i\kern-.025em b}\kern-.08em
    T\kern-.1667em\lower.7ex\hbox{E}\kern-.125emX}}
\begin{document}


\title{School Virus Infection Simulator for Customizing School Schedules During COVID-19*\\
{\footnotesize \textsuperscript{*}}
\thanks{https://github.com/satoshi-takahashi-lab/school-virus-infection-simulator}
}

\author{\IEEEauthorblockN{1\textsuperscript{st} Satoshi Takahashi}
\IEEEauthorblockA{\textit{College of Science and Engineering} \\
\textit{Kanto Gakuin University}\\
Yokohama, Japan \\
0000-0002-1067-6704}
\and
\IEEEauthorblockN{2\textsuperscript{nd} Masaki Kitazawa}
\IEEEauthorblockA{
\textit{Kitazawa Tech}\\
Fujisawa, Japan}
\IEEEauthorblockA{\textit{Graduate School of Artificial Intelligence and Science} \\
\textit{Rikkyo University}\\
Tokyo, Japan \\
0000-0002-6352-0164}
\and
\IEEEauthorblockN{3\textsuperscript{rd} Atsushi Yoshikawa}
\IEEEauthorblockA{\textit{School of Computing} \\
\textit{Tokyo Institute of Technology}\\
Yokohama, Japan}
\IEEEauthorblockA{\textit{Graduate School of Artificial Intelligence and Science} \\
\textit{Rikkyo University}\\
Tokyo, Japan \\
0000-0002-6352-0164}
}

\maketitle

\begin{abstract}
Even in the midst of the Coronavirus (the COVID-19) pandemic, schools strive to provide consistent education to their students.
Teachers and education policymakers are seeking ways to re-open schools, as it is necessary for community and economic development.
However, in the light of the pandemic, schools require customized schedules that can address the health concerns, and safety of the students in respect with classroom sizes, air conditioning equipment, and classroom systems (self-contained or departmentalized).
To solve this issue, we developed the School-Virus-Infection-Simulator (SVIS) for teachers and education policymakers.
The SVIS takes into account the students’ lesson schedules, classroom volume, air circulation rates in the classrooms, and infectability of the students, and simulates the spread of infection at a school.
Thus, teachers and education policymakers can simulate how their school schedules can impact current health concerns.
We then demonstrate the impact of several school schedules in self-contained and departmentalized classrooms, and evaluate them in terms of the maximum number of students infected simultaneously, and the percentage of face-to-face lessons.
The results show that increasing the classroom ventilation rate is effective. However, the impact is not as stable as customizing school schedules.
In addition, school schedules can impact the maximum number of students infected simultaneously differently, depending on whether classrooms are self-contained or departmentalized.
It was found that a school schedule, each student group goes to school for one week and takes days off for three week alternately, had a higher maximum number of students infected simultaneously, compared to schedules with a higher percentage of face-to-face lessons.
The SVIS and the simulation results can help teachers and education policymakers plan school schedules appropriately, in order to reduce the maximum number of students infected simultaneously, while also maintaining a certain percentage of face-to-face lessons.
\end{abstract}


\section{Introduction}
During COVID-19, schools have continuously strived to provide consistent education.
Teachers and education policymakers are seeking ways to re-open schools, which is necessary for community and economic development.

The World Health Organization (WHO) states that schools should assess several elements when deciding to re-open for students.
For instance, the epidemiology of COVID-19 at the local level, the benefits and risks to children and staff, and the capacity of schools/educational institutions to operate safely~\cite{WHO20Schools}.
The benefits of re-opening school include not only educational effects, but also social and psychological well-being of students and staff, essential services, access to nutrition, child welfare, and freedom for parents to work~\cite{WHO20Schools}.

To re-open safely, WHO recommends several measures: wearing a mask,
ensuring adequate air supply in classrooms, and maintaining physical distance between students~\cite{WHO20Schools}.
However, the actual situation in the schools differ.
Some schools, for example, schools in developing or cold counties, do not have the budget to improve the total airflow supply.
Moreover, not all schools have sufficient classrooms, or empty classrooms to maintain physical distance.

Many schools adapt cohorting, and customize school schedules~\cite{US20,Children21}.
These include creating small groups of staff and teachers.
The schools reassemble their schedules, divide students into several groups, conduct face-to-face lessons for one group, give homework to the other groups, and conduct online classes.
This methodology allows students to maintain physical distance, and keep the air clean, without additional classrooms or air conditioners.

UNESCO suggests four types of school schedules, their risk of infection, and the pros and cons of education effects, school management, parents, and so on  ~\cite{UNESCO20}.
According to the schedule provided by UNESCO students are divided into two groups and each group is then scheduled to go to school alternately for half a day, one day, two days, and one week; it further states the ranking of the four schedules by infection risks. However, school schedules can be customized in many ways, and their impact on education, and effectiveness in reducing infectious risks are different.
Teachers and education policymakers have to choose a suitable school schedule from a variety of plans without the complete knowledge of the infection risks of the program.
For example, there are representative school schedules which include self-contained and departmentalized classrooms; dividing the students into two groups would not be enough when the total number of students is large.
Hence, teachers and education policymakers cannot judge the infections risk based on the UNESCO’s ranking.

We propose the School-Virus-Infection-Simulator (SVIS) to simulate the spread of virus infection at a school according to the school schedule.
The SVIS can simulate airborne infection in several school situations, considering the number of students and classrooms, and the performance of air conditioners.
It can help teachers and education policymakers customize school schedules with respect to the infection risk, and the percentage of face-to-face lessons.
In this study, education policymakers mean staff members who are responsible for school scheduling; not for the social economy.

In this paper, we introduce the SVIS and demonstrate that it can evaluate the impact of several school schedules of self-contained and departmentalized classrooms from the viewpoint of the maximum number of students infected simultaneously and the percentage of face-to-face lessons.

\section{Related Work}
Several studies have modeled and simulated the spread of COVID-19 that may be caused by the re-opening of schools.
Their targets are classified, roughly, into two categories: outside and inside schools.
In addition, several school scheduling models are proposed.
This study focuses on a representative school scheduling type with self-contained and departmentalized classrooms.

\subsection{Outside School Model}
Cruz~\cite{Cruz21} simulated school re-opening strategies in the S\~{a}o Paulo Metropolitan Area.
These included both scenarios: re-opening schools with all the students at once, following the S\~{a}o Paulo government plan, and re-opening only when a vaccine becomes available.
He used a stochastic compartmental model that included a heterogeneous and dynamic network.
Gharakhanlou~\cite{Gharakhanlou20} simulated the spatio-temporal outbreak of COVID-19 with an agent-based model.
He investigated the impact of various strategies of school and educational center closure, with special attention to social distancing and office closures, on the COVID-19 outbreak in Urmia city, Iran.
Lee ~\cite{Lee21} simulated the outbreaks in the greater Seattle area, and evaluated the effect of the combination of non-pharmaceutical interventions, such as social distancing, use of face mask, school closure, testing, and contact tracing.
Kim ~\cite{Kim2020-mm} modeled the COVID-19 transmission dynamic in Korea with the susceptible-exposed-infectious-recovered model.
His model considered two age groups, children and adults, because their social behaviors are different. He estimated the effect of delay in school re-opening in Korea.
Chin~\cite{Children21} developed a national- and county- level simulation model, considering school closures and unmet childcare needs in the US.
He estimated the projected rate of unmet childcare needs for healthcare worker households.

\subsection{Inside School Model}
Zafarnejad~\cite{Zafarnejad21} assessed school policy actions for COVID-19 using an agent-based simulation.
He modeled a classroom with two-dimensional tiles, and simulated the spread of COVID-19 quanta in a closed classroom environment.
He compared the infection risks among several non-pharmaceutical interventions, including class schedules, social distancing, ventilation, air filtration, surveillance testing, and contact tracing.
Brom~\cite{Brom21} modeled the interactions among pupils and teachers in Prague, Czechia as a multi-graph structure with an agent-based simulation.
He investigated the impact of several school schedule types, and antigen and PCR test schedule types on reducing the spread.
Ghaffarzadegan~\cite{Ghaffarzadegan21} developed a hypothetical university model of 25,000 students, and 3,000 faculty/staff in a U.S. college town with a mathematical and compartmental model.
He simulated several combinations of policies, and evaluated the impact of COVID-19.
The policies tested included proactive and quick testing, high mask adoption, better risk communication with students, and remote work for high-risk individuals.
Bilinski~\cite{Bilinski21} developed an agent-based network model to simulate transmission in elementary and middle school communities. He built three screening strategies - weekly or bi-weekly screening of all students and teachers, and a 24-hour test turnaround time.
He then compared these screening scenarios with three school scenarios - five-day in-person attendance, a hybrid model in which 50\% students attended class on Monday and Tuesday, and the other 50\% on Thursday and Friday, and complete remote learning.
McPeck~\cite{McPeck21} built an agent-based model in which students moved, and interacted on a dormitory floor of Eastern Washington University. He simulated multiple scenarios with a combination of vaccination and masking rates.

\subsection{School Scheduling}
\begin{figure}[tb]
  \centering
  \includegraphics[width=1.0\linewidth]{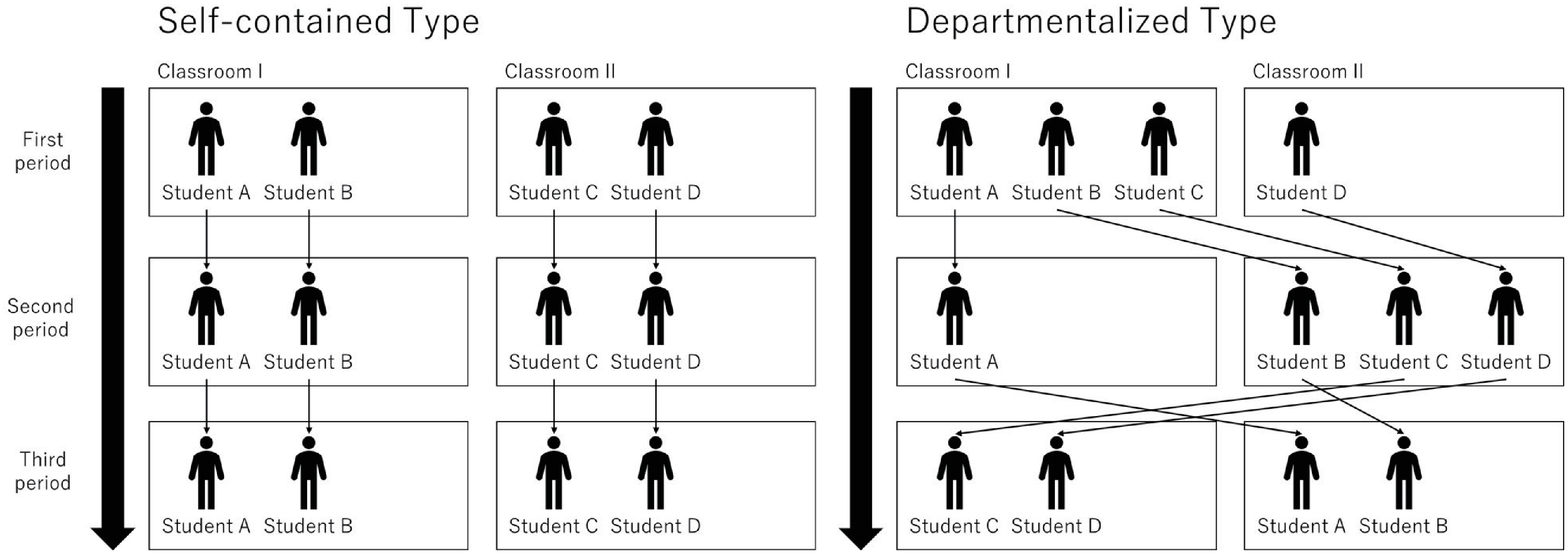}
  \caption{Classroom type}
  \label{fig:Classroomtype}
\end{figure}

Schools consider teachers’ cost and students’ learning effect when making school schedules. One such representative school scheduling type includes self-contained, and departmentalized classrooms (Fig.~\ref{fig:Classroomtype}).
Students of self-contained classrooms took the same lesson in the same classroom. This could be regarded as a type of cohorting.
However, students in departmentalized classrooms took individual lessons in different classrooms.
Self-contained classrooms required teachers to specialize in multiple subjects ~\cite{ERS20}, whereas teachers of departmentalized classrooms focused on only one subject.
Students studying in lower grades, or those with low-incidence disabilities, need learning support.
The teachers of self-contained classrooms can observe these students continuously, whereas, the teachers of departmentalized classrooms can observe them once a day, or once in two days.
Therefore, schools tend to adopt self-contained classrooms for lower grades, and departmentalized classrooms for higher grades.
However, some high schools have recently mandated self-contained classrooms to reduce student interactions~\cite{CTC21Educator}.

\subsection{Summary}
Our simulation model was classified into the inside school model.
Previous studies assumed that all students' behavior was the same, and targeted only one type of school system.
Despite this, we focused on the varied effect of school schedules on COVID-19 on different school types.

Previous studies, that considered external factors, did not simulate and analyze the details of internal school factors.
However, UNESCO pointed out that the effect of internal school factors was very important~\cite{UNESCO20}.
Tupper ~\cite{Tupper2021-zx} and Leng ~\cite{Leng2021-aw} pointed out the risk of COVID-19 clusters inside schools.
We focused on the inside of schools.
We assumed that the external factors could be replaced as the number of students infected outside a school per unit time.
In the experiment sections of this study, we simulated the effect of one infected student; The student is first infected person in the school and no students have an antibody against the COVID-19.
It is noted that our model can simulate scenarios where multiple students get infected at various times.

Furthermore, schools are caught in a dilemma between self-contained and departmentalized classrooms, considering teachers’ cost, students’ learning effect, and reduction in infection.
Self-contained and departmentalized classrooms would react differently to school schedule interventions.
In this study, we shed light on the infection phenomena of self-contained and departmentalized classrooms.

\section{School Virus Infection Simulator}
The SVIS is a tool used to simulate the spread of virus infection at a school.
It simulates students’ behavior based on their lesson schedule, classroom volume, classroom air change rate, and infectability of the virus.
The SVIS is based on an extended version of the susceptible-exposed-infectious-removed model, and agent-based model~\cite{He2020seir,Annas20stability,Mwalili20seir,Yang20modified,Abdou12designing,Cuevas20agent}.

\subsection{School Scheduling Model}
SVIS is based on an agent-based model where every student (agent) attends lessons based on their individual schedules; the SVIS can simulate a wide variety of school scheduling.
In one instance, all students come to school for two to three weeks, after which they stay at home for a week.
WHO states that ``the average time from exposure to COVID-19 to the onset of the symptoms is 5-6 days'' ~\cite{WHO20}.
Hence, when students are infected, their symptoms would appear during the home week, and they can take the following weeks off.
This decreases the risk of infection.
In another instance, students are divided into four groups.
These groups come to school during alternate weeks so that they can interact with more peers.
Teachers and education policymakers have to choose a school schedule from the available options without prior knowledge of the infection risk of the schedule.

\begin{figure}[tb]
  \centering
  \includegraphics[width=1.0\linewidth]{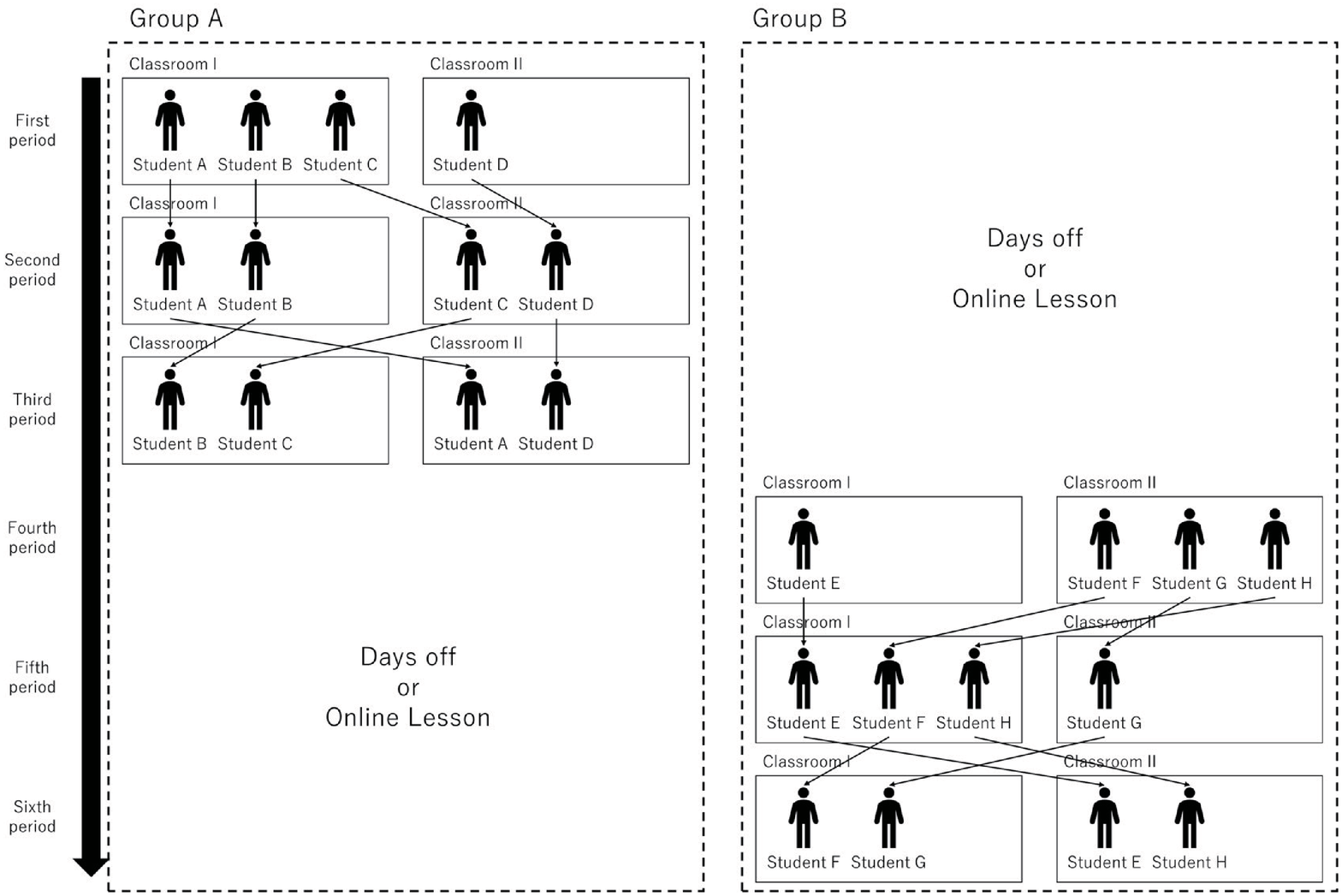}
  \caption{School scheduling model example}
  \label{fig:SchoolSchedulingModel}
\end{figure}

Fig.~\ref{fig:SchoolSchedulingModel} shows an example of school scheduling.
Students are divided into two groups - Group A and Group B.
Each group goes to school for half a day, alternatively.
This schedule corresponds to UNESCO’s school schedule option 1~\cite{UNESCO20}.
In addition, the classroom type is departmentalized.
For example, student A takes first and second period lesson in classroom I, third period lesson in classroom II, and then takes the day off for the fourth, fifth, and sixth periods.

\subsection{Infection Model}
\begin{figure}[tb]
  \centering
  \includegraphics[width=1.0\linewidth]{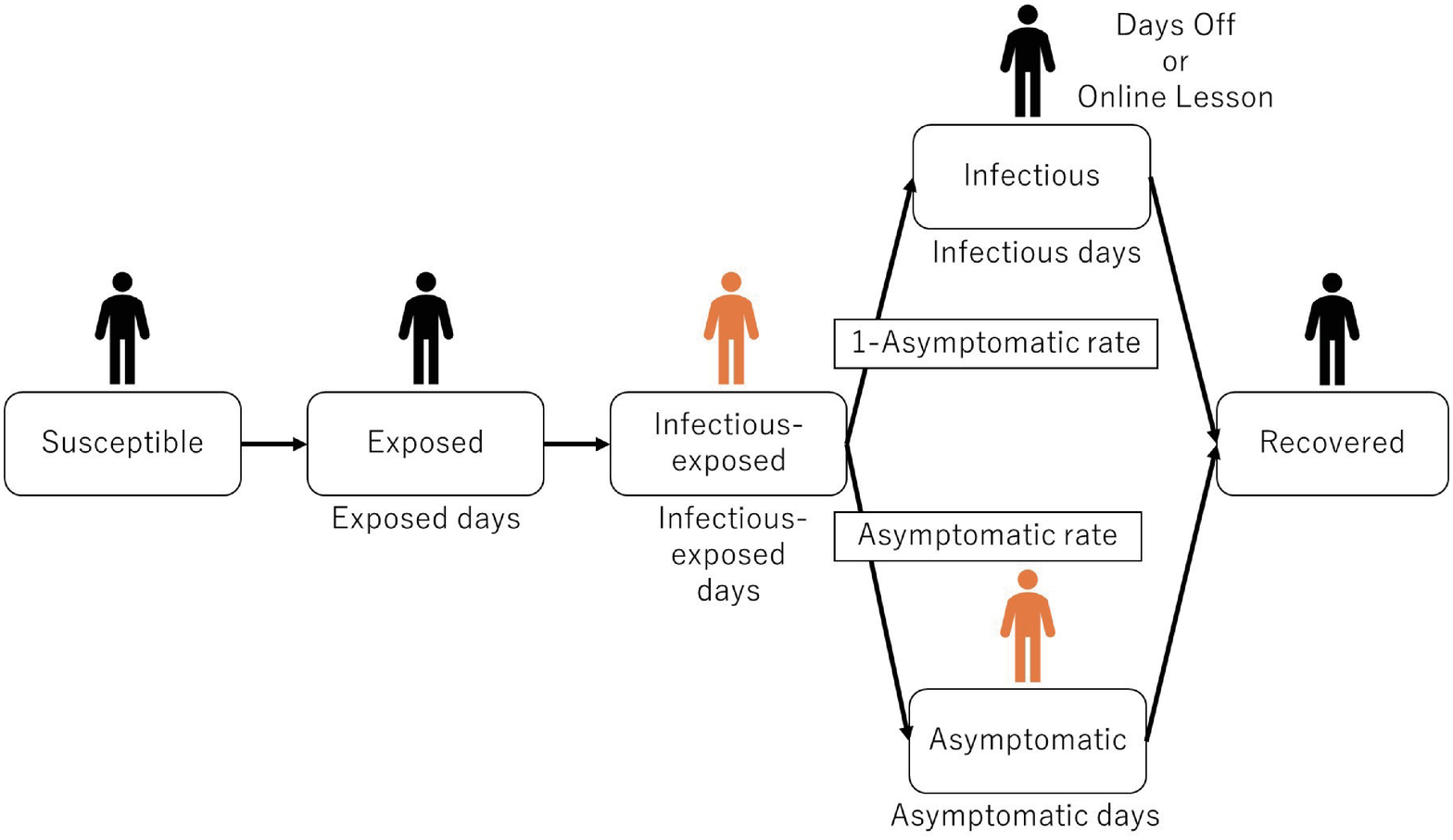}
  \caption{Infection model}
  \label{fig:InfectionModel}
\end{figure}

Students’ status consists of susceptible, exposed, infectious-exposed,  infectious, asymptomatic, and recovered (Fig.~\ref{fig:InfectionModel}).
Exposed individuals do not pose the risk of infecting others while infections-exposed can infect other people.
Susceptible students become exposed according to the infection probability. Exposed students become infectious-exposed after a certain period.
Infectious-exposed students become infectious or asymptomatic, probabilistically, after a certain period.
Infectious and asymptomatic students recover after a specific period.
Infectious-exposed and asymptomatic students infect susceptible students according to infection probability.
Infectious components are the students who develop symptoms, and take the day off or take online lesson.
Asymptomatic students are those who do not develop symptoms, and continue attending school.

Infection probability is calculated based on Dai’s extended edition of Wells-Riley equation~\cite{Dai20,Riley78,Wells55}.

\begin{eqnarray}\label{eq:p}
P = \frac{C}{S} = 1-e^{-Iqpt(1-n_{I})(1-n_{S})/Q}
\end{eqnarray}

$P$ is the probability of infection, $C$ is the number of new infection cases, $S$ is the number of susceptible people, $I$ is the number of infectors (infectious-exposed and asymptomatic students), $q$ is the quanta generation rate, $p$ is the pulmonary ventilation rate of a student, $t$ is the lesson time interval, $n_{I}$ is the exhalation filtration efficiency, $n_{S}$ is the respiration filtration efficiency, and $Q$ is the classroom ventilation rate with clean air.

\begin{figure}[tb]
  \centering
  \includegraphics[width=1.0\linewidth]{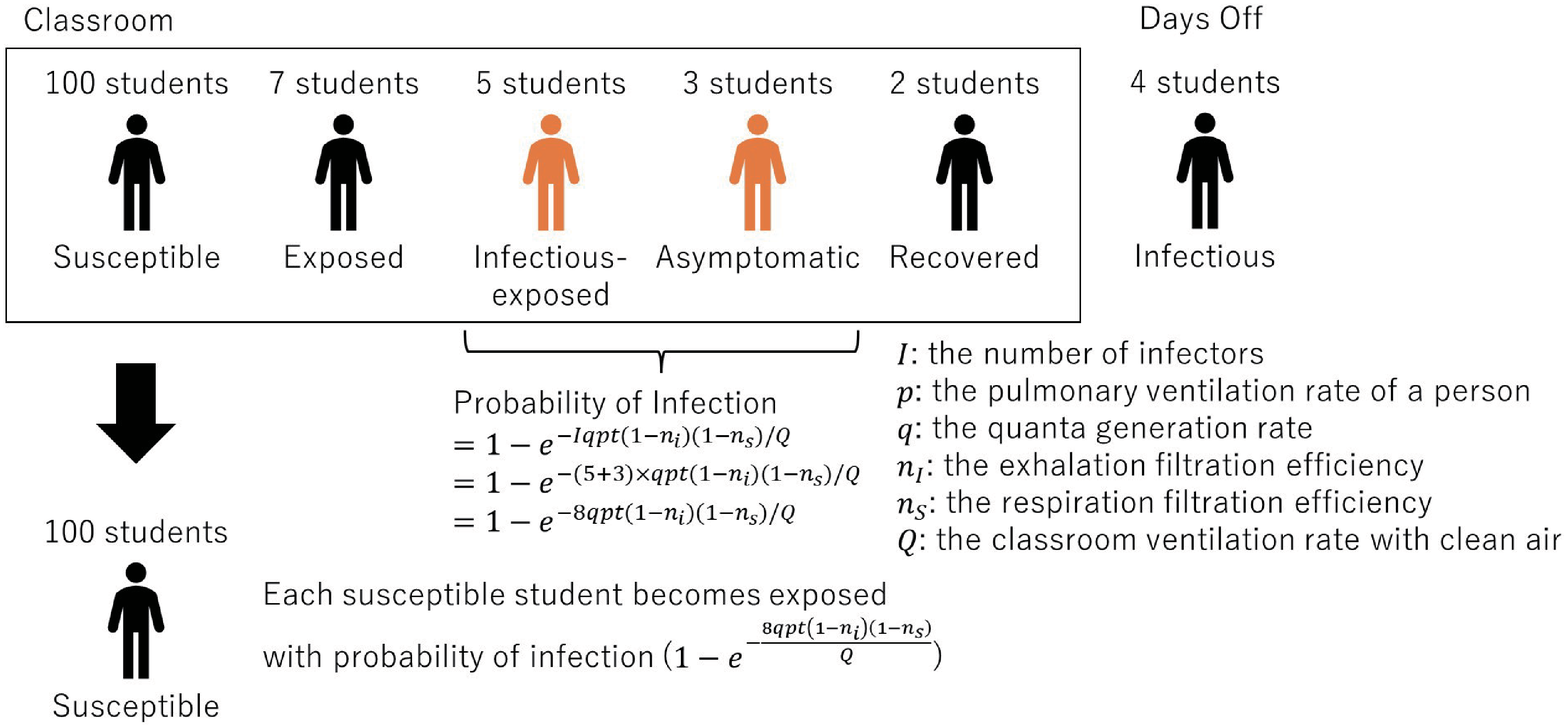}
  \caption{Infection example}
  \label{fig:InfectionExample}
\end{figure}

A student takes lessons based on their schedule, and the infection probability is calculated using (\ref{eq:p}).
Fig.~\ref{fig:InfectionExample} shows the calculation of new infection cases.
The Wells-Riley equation is generally used to calculate the basic reproduction number of the infection ($C/I$).
Therefore, we should calculate the new infection cases as the number of susceptible students multiplied by~(\ref{eq:p}).
However, the number of students in the classroom is often less than 100, and the probability of infection is generally less than 0.01.
Thus, the new infection cases become less than one student. Therefore, we used ~(\ref{eq:p}) as the probability of infection for each susceptible student.
The expectation value of the new infection cases fits the basic reproduction number, using multiple simulations.
In Fig.~\ref{fig:InfectionExample}, 100 susceptible students, five infectious-exposed students, and three asymptomatic students take lessons in the same classroom.
Then, $I$ in~(\ref{eq:p}) becomes eight, and each susceptible student becomes exposed with the probability of infection $(1-e^{-8qpt(1-n_{I})(1-n_{S})/Q})$.

\subsection{Summary}

The SVIS enables teachers and education policymakers to simulate the effects of their school policies in specific situations.
Every school’s policy might be different.
The lesson schedule, and classroom volume vary according to school, country, and state.
Moreover, the budget is different.
Even if teachers and education policymakers in developing countries know that a high ventilation rate with clean air is effective, they would not have enough funding to change the air conditioners.
The SVIS can consider these problems and help them plan lesson schedules to reduce infection probability without replacing air conditioners.

\section{Experiment Design}
First, we show the effects of changing classroom volumes, and classroom air change rates during COVID-19.
Next, we demonstrate the impact of several school schedules in self-contained, and departmentalized classrooms, and evaluate those schedules for the maximum number of students infected simultaneously, and the percentage of face-to-face lessons.
The former is an essential indicator for controlling COVID-19 because hospitals must accommodate isolated beds, and medical equipment (e.g., extracorporeal membrane oxygenation) for infected people~\cite{Barbaro20}.
The percentage of face-to-face lessons is also essential for understanding educational effect and students’ motivation, and building classroom community ~\cite{WHO20Schools}.

\subsection{Basic Parameters}
Table~\ref{tab:BasicParameters} lists the basic parameters in all the experiments.
Buonanno estimated the pulmonary ventilation rate of sedentary activity as 0.54 (m3/h)~\cite{Buonanno20,Adams93}.
Dai calculated the quanta generation rate of COVID-19 as 14-48 in 2020~\cite{Dai20}.
We adopt 48 as $q$ because the Delta variant is spreading worldwide, and the infectability is estimated to be stronger than the original~\cite{CDC21Delta}.
Dai estimated the exhalation filtration efficiency, and the respiration filtration efficiency as 0.5 when all students wear a mask~\cite{Dai20}.
The CDC says, ``isolation, and precautions can be discontinued 10 days after symptom onset, and after resolution of fever for at least 24 hours, and improvement of other symptoms''~\cite{CDC21Ending}.
Thus, we roughly estimated 14 as infected days.
Dai estimated that infectability is at a peek two days before and one day after symptom onset.
WHO states, ``the time from exposure to COVID-19 to the moment when symptoms start to show, on average, is 5 to 6 days.''
Hence, we adopted three days, and two days as exposed days, and infectious-exposed days, respectively~\cite{He20,WHO20}.
Bullard estimated that SARS-CoV-2-infected Vero cell infectivity is observed only eight days after symptom onset~\cite{Bullard20}.
Hence, we adopted eight days as asymptomatic days.
Also, we considered the CDC estimate of asymptomatic infection percentage as 30\% (the current best)~\cite{CDC21Planning}.

\begin{table}[bt]
  \centering
  \caption{Basic parameters}
  \label{tab:BasicParameters}
  \begin{tabular}{lc}\toprule
    \textit{Item} & \textit{Value}\\ \midrule
    $p$: the pulmonary ventilation rate of a person & 0.54(m3/h) \\
    $q$: the quanta generation rate & 48 \\
    $n_{I}$: the exhalation filtration efficiency & 0.5 \\
    $n_{S}$: the respiration filtration efficiency & 0.5 \\
    Asymptomatic rate & 0.3 \\
    Exposed days & Three days \\
    Infectious-exposed days & Two days \\
    Infectious days & 14 days \\
    Asymptomatic days & Eight days \\ \bottomrule
  \end{tabular}
\end{table}

\subsection{School Schedules}
We designed several school schedules (Tables~\ref{tab:MiddleschoolsandHighschoolsParameters}, ~\ref{tab:SchoolSchedules1}, and ~\ref{tab:SchoolSchedules2}).
A school chooses self-contained or departmentalized classroom.
A school divides students into several groups; some groups take face-to-face lessons; the other groups do homework or take online classes for several periods.
The groups take face-to-face lessons alternatively.

\begin{figure}[bt]
  \centering
  \includegraphics[width=1.0\linewidth]{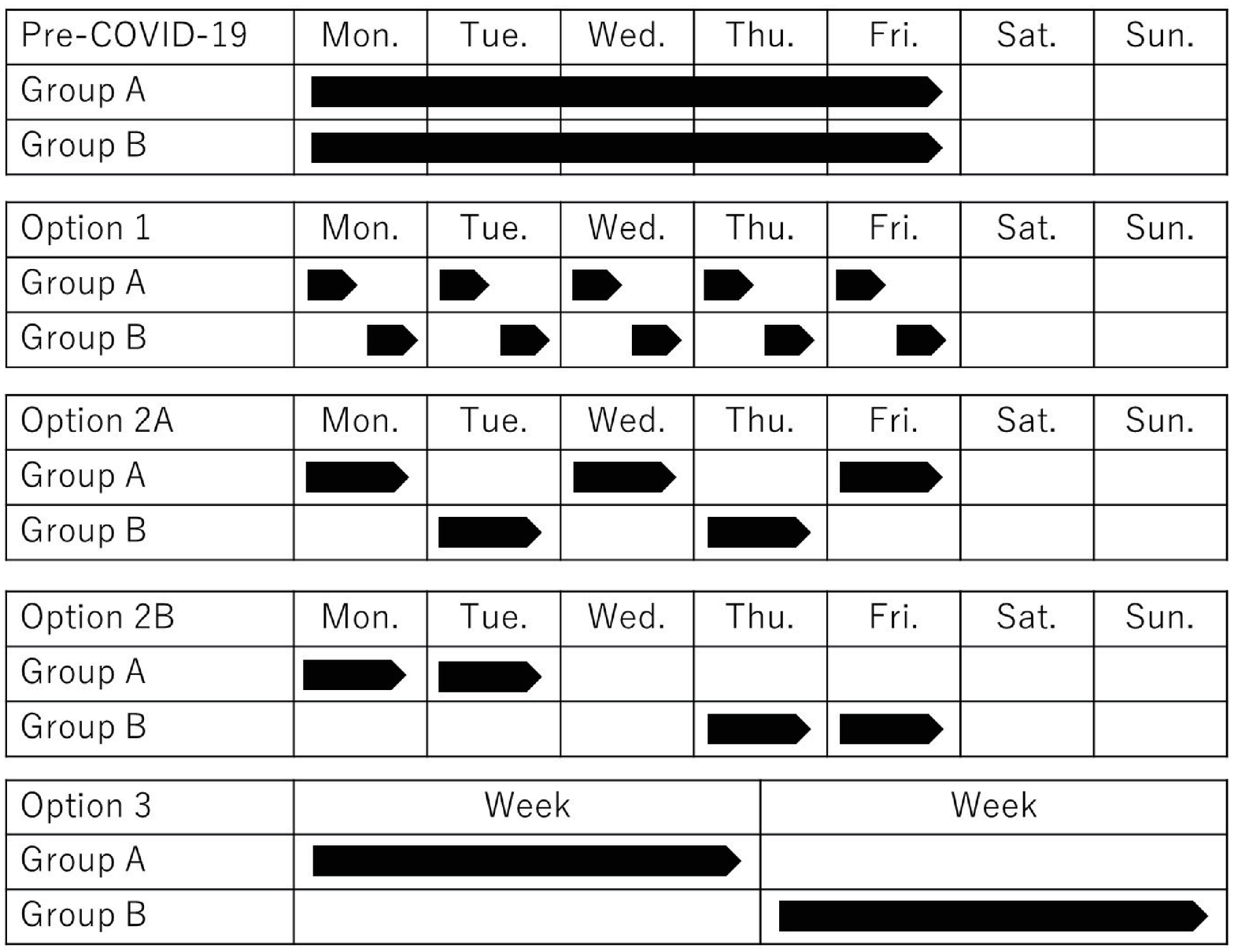}
  \caption{UNESCO school schedule~\cite{UNESCO20}}
  \label{fig:UNESCOSchoolSchedule}
\end{figure}

\begin{table*}[!htb]
\centering
\begin{threeparttable}
\centering
  \caption{Self-contained classrooms and departmentalized classrooms parameters}
  \label{tab:MiddleschoolsandHighschoolsParameters}
\begin{tabular}{lcl} \toprule
Item                                                                                                                                           & Self-contained & \multicolumn{1}{c}{Departmentalized} \\ \midrule
Classroom volume                                                                                                                               & \multicolumn{2}{c}{150 m3}                            \\
Classroom air change rate \tnote{a}                                                                                                                     & \multicolumn{2}{c}{Three time/h}                      \\
Classroom ventilation rate with clean air & \multicolumn{2}{c}{450 m3/h}                          \\
Lesson time                                                                                                                                    & \multicolumn{2}{c}{50 min}                            \\
Total number of students                                                                                                                       & 24 students      & 480 students                       \\
Number of classrooms per lesson                                                                                                                & One classroom    & 20 classrooms                      \\
Lesson weeks  \tnote{b}                       & \multicolumn{2}{c}{12 weeks}                          \\
Number of lessons per day                                                                                                                      & \multicolumn{2}{c}{7 lessons}  \\ \bottomrule                   
\end{tabular}
\begin{tablenotes}
\item[a] Classroom volume multiplied by classroom air change rate
\item[b] Five days are lesson days and two days are days off in a week
\end{tablenotes}
\end{threeparttable}
\end{table*}

\begin{table*}[!htb]
\centering
\caption{School schedules}
\label{tab:SchoolSchedules1}
\begin{tabular}{lcccccc} \toprule
                                                                                                                              & \multicolumn{6}{c}{School Schedule Name}                                                                                                                                                                                                             \\
                                                                                                                              & \begin{tabular}[c]{@{}c@{}}Type (i)\\Pre-COVID-19\end{tabular}   & Type (ii) & Type (iii)                                    & Type (iv)                                   & Type (v)                               & \begin{tabular}[c]{@{}c@{}}Type (vi)\\UNESCO 3\end{tabular}                                \\ \midrule  \midrule
Schedule Category                                                                                                        & Basic          & Shortened         & Weeks                                                 & Weeks                                                 & Weeks                                              & Weeks                                              \\ \midrule
Number of Groups                                                                                                     & 1              & 1                  & 2                                                     & 4                                                     & 1                                                  & 2                                                  \\ \midrule
\begin{tabular}[c]{@{}l@{}}Number of Face-to-Face Lesson Groups\\ at One Time\end{tabular}                           & 1              & 1                  & 1                                                     & 3                                                     & 1                                                  & 1                                                  \\ \midrule
\begin{tabular}[c]{@{}l@{}}A Continuous Period of Face-to-Face Lesson\\ for Each Group\end{tabular}                  & Always         & Always             & \begin{tabular}[c]{@{}c@{}}Three\\ weeks\end{tabular} & \begin{tabular}[c]{@{}c@{}}Three\\ weeks\end{tabular} & \begin{tabular}[c]{@{}c@{}}One\\ week\end{tabular} & \begin{tabular}[c]{@{}c@{}}One\\ week\end{tabular} \\ \midrule
\begin{tabular}[c]{@{}l@{}}A Continuous Period of Not Going to a School\\ for Each Group\end{tabular}                & No             & No                 & \begin{tabular}[c]{@{}c@{}}One\\ week\end{tabular}    & \begin{tabular}[c]{@{}c@{}}One\\ week\end{tabular}    & \begin{tabular}[c]{@{}c@{}}One\\ week\end{tabular} & \begin{tabular}[c]{@{}c@{}}One\\ week\end{tabular} \\ \midrule
Percentage of Face-to-Face Lessons                                                                                   & 100\%          & 100\%              & 75\%                                                  & 75\%                                                  & 50\%                                               & 50\%                                               \\ \midrule
\begin{tabular}[c]{@{}l@{}}Percentage of Class Members That the Students Meet\\in Face-To-Face Lessons\end{tabular} & 100\%          & 100\%              & 100\%                                                 & 100\%                                                 & 100\%                                              & 50\%   \\ \bottomrule                                           
\end{tabular}
\end{table*}

\begin{table*}[!htb]
\centering
\caption{School schedules}
\label{tab:SchoolSchedules2}
\begin{tabular}{lcccccc} \toprule
                                                                                                                              & \multicolumn{6}{c}{School Schedule Name}                                                                                                                                                                                                                                                                                   \\
                                                                                                                              & \begin{tabular}[c]{@{}c@{}}Type (vii)\\UNESCO 2B\end{tabular}                                                         & \begin{tabular}[c]{@{}c@{}}Type (viii)\\UNESCO 2A\end{tabular}                                & \begin{tabular}[c]{@{}c@{}}Type (ix)\\UNESCO 1\end{tabular}
 & Type (x)                                            & Type (xi)                                   & Type (xii)                                  \\ \midrule  \midrule
Schedule Type                                                                                                        & Days                                                                      & Days                                              & Days                & Weeks                                                         & Weeks                                                 & Weeks                                                 \\ \midrule
Number of Groups                                                                                                     & 2                                                                         & 2                                                 & 2                   & 4                                                             & 1                                                     & 4                                                     \\ \midrule
\begin{tabular}[c]{@{}l@{}}Number of Face-to-Face Lesson Groups\\ at One Time\end{tabular}                          & 1                                                                         & 1                                                 & 1                   & 2                                                             & 1                                                     & 1                                                     \\ \midrule
\begin{tabular}[c]{@{}l@{}}A Continuous Period of Face-to-face Lesson\\ for Each Group\end{tabular}                  & \begin{tabular}[c]{@{}c@{}}Two and\\a half days\end{tabular} & \begin{tabular}[c]{@{}c@{}}One\\ day\end{tabular} & A half-day          & \begin{tabular}[c]{@{}c@{}}Depends\\ on\\ groups\end{tabular} & \begin{tabular}[c]{@{}c@{}}One\\ week\end{tabular}    & \begin{tabular}[c]{@{}c@{}}One\\ week\end{tabular}    \\ \midrule
\begin{tabular}[c]{@{}l@{}}A Continuous Period of Not Going to a School\\ for Each Group\end{tabular}                & \begin{tabular}[c]{@{}c@{}}Two and\\ a half days\end{tabular} & \begin{tabular}[c]{@{}c@{}}One\\ day\end{tabular} & A half-day          & \begin{tabular}[c]{@{}c@{}}Depends\\ on\\ groups\end{tabular} & \begin{tabular}[c]{@{}c@{}}Three\\ weeks\end{tabular} & \begin{tabular}[c]{@{}c@{}}Three\\ weeks\end{tabular} \\ \midrule
Percentage of Face-to-Face Lessons                                                                                   & 50\%                                                                      & 50\%                                              & 50\%                & 50\%                                                          & 25\%                                                  & 25\%                                                  \\ \midrule
\begin{tabular}[c]{@{}l@{}}Percentage of Class Members That the Students Meet\\in Face-To-Face Lessons\end{tabular} & 50\%                                                                      & 50\%                                              & 50\%                & 100\%                                                         & 100\%                                                 & 25\%   \\ \bottomrule                                                     
\end{tabular}
\end{table*}

\begin{table*}[!htb]
\centering
\caption{Number of simulations of self-contained classroom}
\label{tab:NumberofMiddleSchoolSimulations}
\begin{tabular}{lcccc}\toprule
\multicolumn{1}{c}{School Schedule Name} & Classrooms Patterns & \begin{tabular}[c]{@{}c@{}}Face-to-Face Lesson\\ Schedule Patterns\end{tabular} & \begin{tabular}[c]{@{}c@{}}Number of Simulations\\ in Each Case\end{tabular} & Total Number of Simulations \\\midrule\midrule
Type (i)                              & 1                   & 1                                                                               & 3,600                                                                        & 3,600                       \\\midrule
Type (ii)                             & 1                   & 1                                                                               & 3,600                                                                        & 3,600                       \\\midrule
Type (iii)                            & 1                   & 1                                                                               & 3,600                                                                        & 3,600                       \\\midrule
Type (iv)                             & 1                   & 18                                                                              & 200                                                                          & 3,600                       \\\midrule
Type (v)                              & 1                   & 1                                                                               & 3,600                                                                        & 3,600                       \\\midrule
Type (vi)                             & 1                   & 1                                                                               & 3,600                                                                        & 3,600                       \\\midrule
Type (vii)                            & 1                   & 1                                                                               & 3,600                                                                        & 3,600                       \\\midrule
Type (viii)                           & 1                   & 1                                                                               & 3,600                                                                        & 3,600                       \\\midrule
Type (ix)                             & 1                   & 1                                                                               & 3,600                                                                        & 3,600                       \\\midrule
Type (x)                              & 1                   & 48                                                                              & 75                                                                           & 3,600                       \\\midrule
Type (xi)                             & 1                   & 1                                                                               & 3,600                                                                        & 3,600                       \\\midrule
Type (xii)                            & 1                   & 6                                                                               & 600                                                                          & 3,600       \\ \bottomrule         
\end{tabular}
\end{table*}

\begin{table*}[!htb]
\centering
\caption{Number of simulations of departmentalized classroom}
\label{tab:NumberofHighSchoolSimulations}
\begin{tabular}{lcccc}\toprule
\multicolumn{1}{c}{School Schedule Name} & Classrooms Patterns & \begin{tabular}[c]{@{}c@{}}Face-to-Face Lesson\\ Schedule Patterns\end{tabular} & \begin{tabular}[c]{@{}c@{}}Number of Simulations\\ in Each Case\end{tabular} & Total Number of Simulations \\\midrule\midrule
Type (i)                              & 20                  & 1                                                                               & 100                                                                          & 2,000                       \\\midrule
Type (ii)                             & 20                  & 1                                                                               & 100                                                                          & 2,000                       \\\midrule
Type (iii)                            & 20                  & 1                                                                               & 100                                                                          & 2,000                       \\\midrule
Type (iv)                             & 20                  & 18                                                                              & 20                                                                          & 7,200                       \\\midrule
Type (v)                              & 20                  & 1                                                                               & 100                                                                           & 2,000                       \\\midrule
Type (vi)                             & 20                  & 1                                                                               & 100                                                                          & 2,000                       \\\midrule
Type (vii)                            & 20                  & 1                                                                               & 100                                                                          & 2,000                       \\\midrule
Type (viii)                           & 20                  & 1                                                                               & 100                                                                          & 2,000                       \\\midrule
Type (ix)                             & 20                  & 1                                                                               & 100                                                                          & 2,000                       \\\midrule
Type (x)                              & 20                  & 48                                                                              & 10                                                                           & 9,600                       \\\midrule
Type (xi)                             & 20                  & 1                                                                               & 100                                                                          & 2,000                       \\\midrule
Type (xii)                            & 20                  & 6                                                                               & 25                                                                           & 3,000     \\ \bottomrule                 
\end{tabular}
\end{table*}

\begin{figure}[bt]
  \centering
  \includegraphics[width=1.0\linewidth]{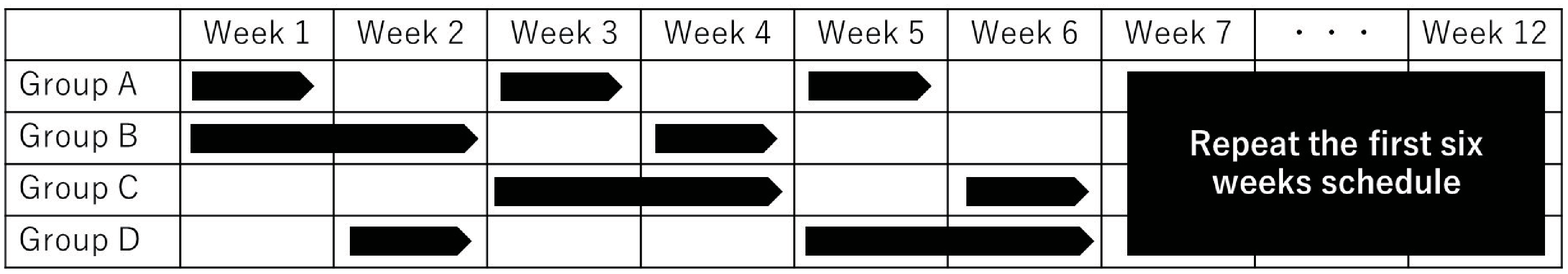}
  \caption{Example of school schedule type (x)}
  \label{fig:pattern x}
\end{figure}

UNESCO proposes four types of school schedules~\cite{UNESCO20} (Fig.~\ref{fig:UNESCOSchoolSchedule}) - Option 1 is an hour-based model, Option 2A and 2B are day-based models, and Option 3 is a week-based model.
UNESCO states that infection risk decreases in this order~\cite{UNESCO20}.
These models have different advantages and disadvantages.
For example, in Option 1 ``students constantly interact with peers, improving their emotional connection.''
However, it is ``logistically demanding for parents as the face-to-face instruction time is short.''
We made twelve school schedule types (Tables ~\ref{tab:NumberofMiddleSchoolSimulations} and ~\ref{tab:NumberofHighSchoolSimulations}).
School schedule type (i) is the basic schedule.
In type (i), five days are active school days and two days are off every week with seven lessons every day from the first day to the fifth day.
School schedule type (ii) is a shortened schedule compared to type (i).
In type (ii), six days are active school days, one day is off every week with six lessons every day from the first day to the fifth day, and five lessons on the sixth day.
The school schedule types (ix), (viii), (vii), and (vi) correspond to UNESCO’s options 1, 2A, 2B, and 3, respectively.
An example of generating a school schedule type (x), and simulating it in departmentalized classrooms is shown below.
\begin{enumerate}[step(1)]
 \item We accumulated students from 1 to 480.
 \item We divided the students into four groups: Group A included students 1 to 120, Group B included students 121 to 240, Group C included students 241 to 360, and Group D included 361 to 480.
 \item We generated six combinations of the two groups, who took face-to-face lessons together. The combinations were (A, B), (A, C), (A, D), (B, C), (B, D) and (C, D).
 \item We assigned the groups for face-to-face lesson schedules for six weeks. We generated their patterns in every combination under two limiting conditions: Group A was permanently set for face-to-face lessons in the first week, and no group took face-to-face lessons for three weeks straight. We found 48 patterns that fulfilled the conditions.
 \item After the first six weeks, the school would repeat the schedule of the first six weeks (Fig.~\ref{fig:pattern x})
 \item We adjusted the number of simulations in each case by considering that the total number of simulations of type (x) was close to the other types’ total number of simulations.
\end{enumerate}

The school schedule type (x) included all the combinations of the two groups, and students could meet all the members in face-to-face lessons.
The percentage of class members that the students met in face-to-face lessons was 100\%.

Every school schedule made student only one in Group A intectious exposed on Day 0 (Monday), and simulated virus infection spread for 12 weeks.

\section{Experiment 1}
\begin{figure}[!t]
  \centering
  \includegraphics[width=\linewidth]{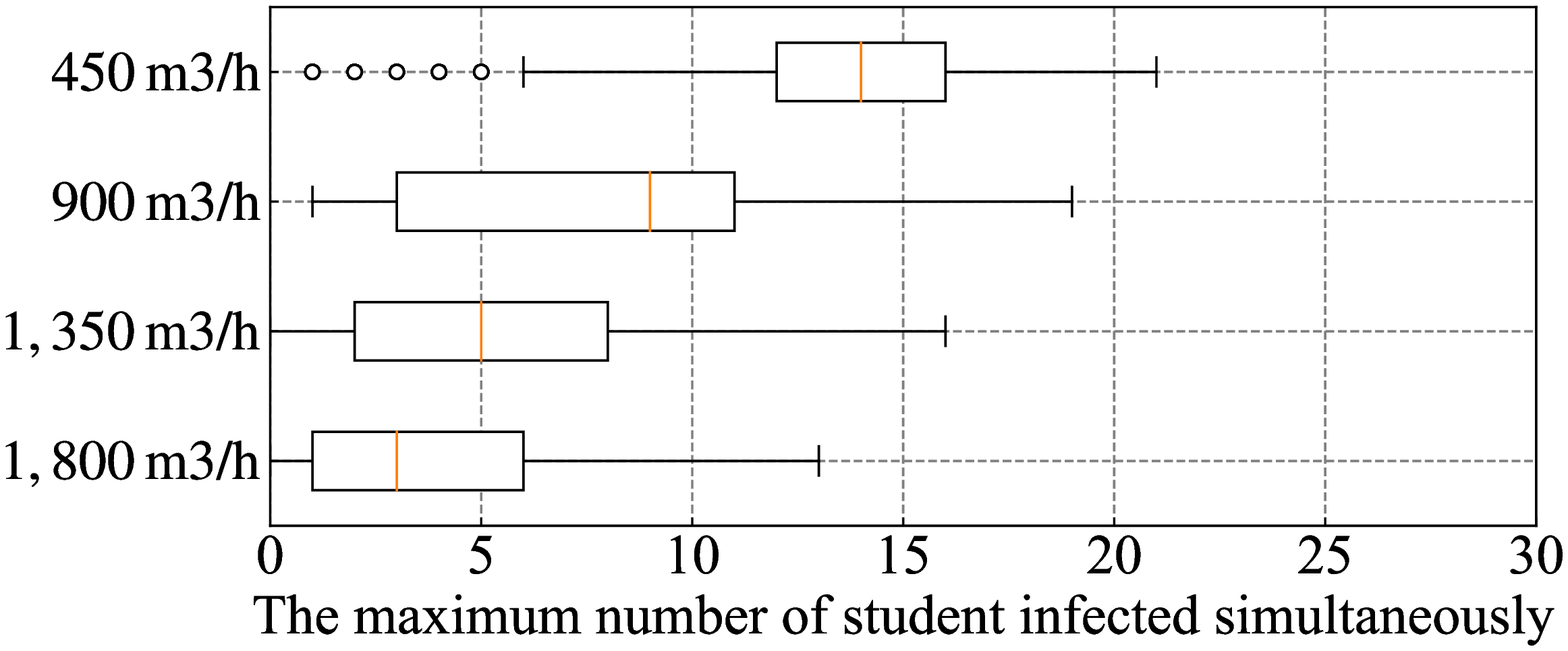}
  \caption{Experiment 1: The maximum number of students infected simultaneously in self-contained classrooms}
  \label{fig:TheMaxNumMiddleSchoolExp1}
  \centering
  \includegraphics[width=\linewidth]{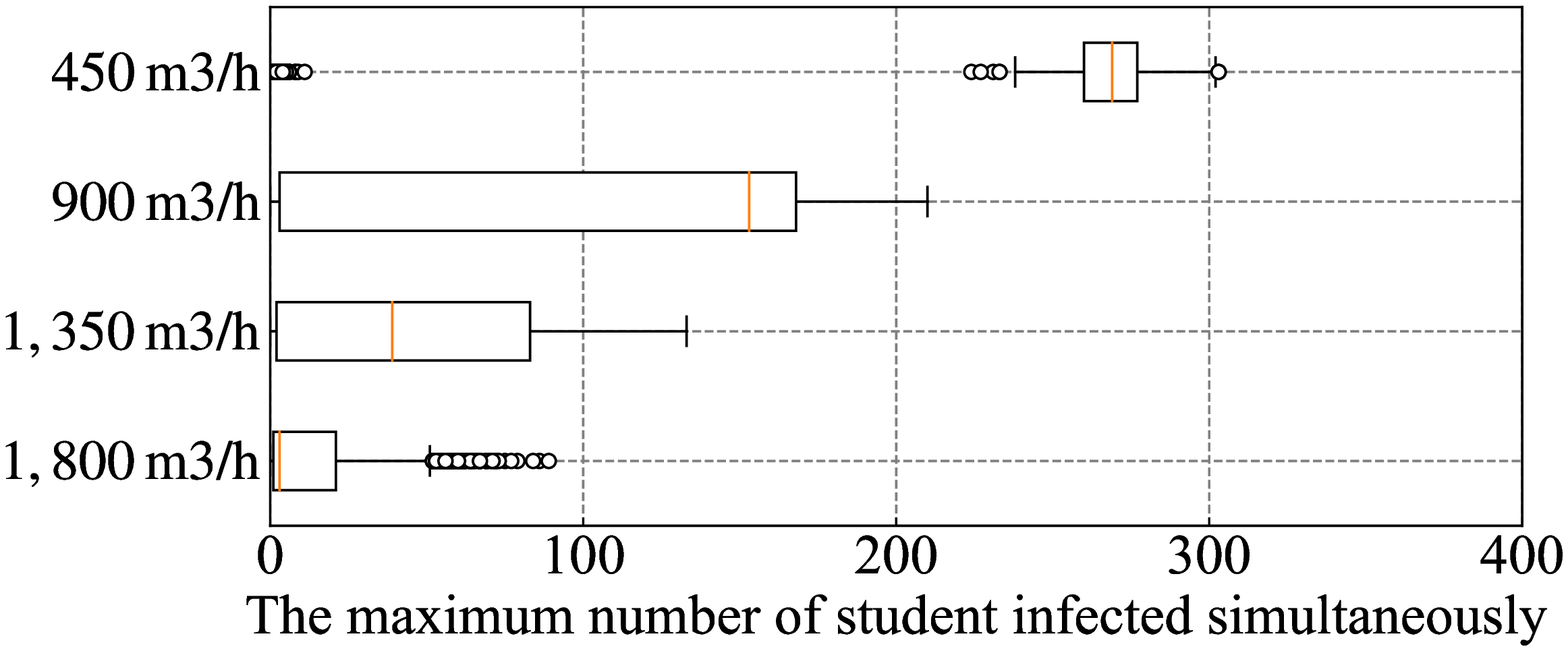}
  \caption{Experiment 1: The maximum number of students infected simultaneously in departmentalized classrooms}
  \label{fig:TheMaxNumHighSchoolExp1}
\end{figure}

To evaluate the effects of classroom volumes, and classroom air change rates, we used type (i) as a basic school schedule, and changed the classroom ventilation rate with clean air (Q in \ref{eq:p}) to 450 m3/h, 900 m3/h, 1,350 m3/h, and 1,800 m3/h.
This experimental design corresponded to a change in the volume or air rate as the basic parameter, doubled, tripled, and quadrupled.
Fig.~\ref{fig:TheMaxNumMiddleSchoolExp1} and ~\ref{fig:TheMaxNumHighSchoolExp1} present the results.
The maximum number of students infected simultaneously decreased, and their variance increased as the classroom ventilation rate increased.
These results show that increasing the classroom ventilation rate is effective in decreasing the spread of COVID-19.

\begin{figure}[!t]
  \centering
  \includegraphics[width=\linewidth]{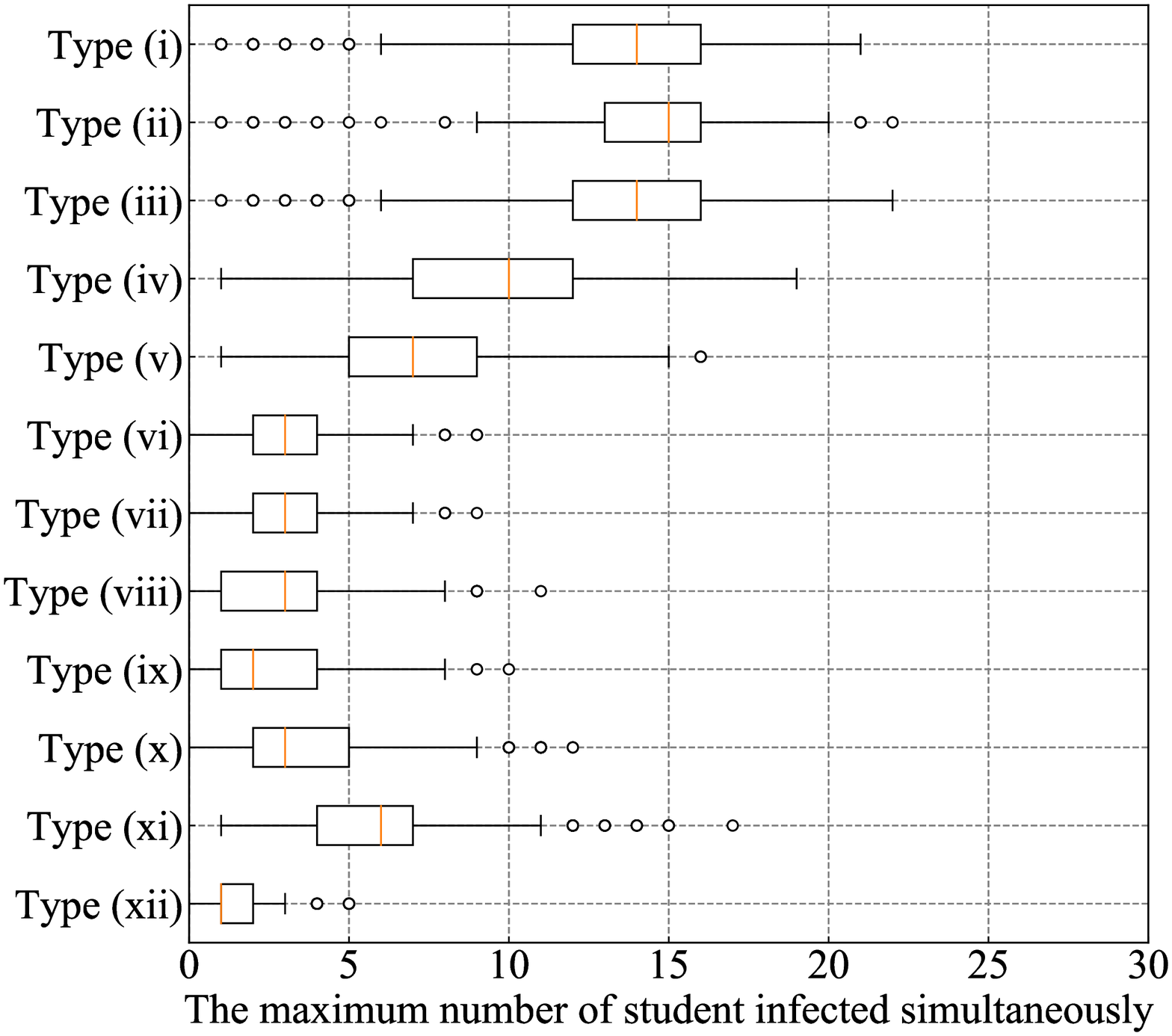}
  \caption{Experiment 2: The maximum number of students infected simultaneously in self-contained classrooms}
  \label{fig:TheMaxNumMiddleSchoolExp2}
  \centering
  \includegraphics[width=\linewidth]{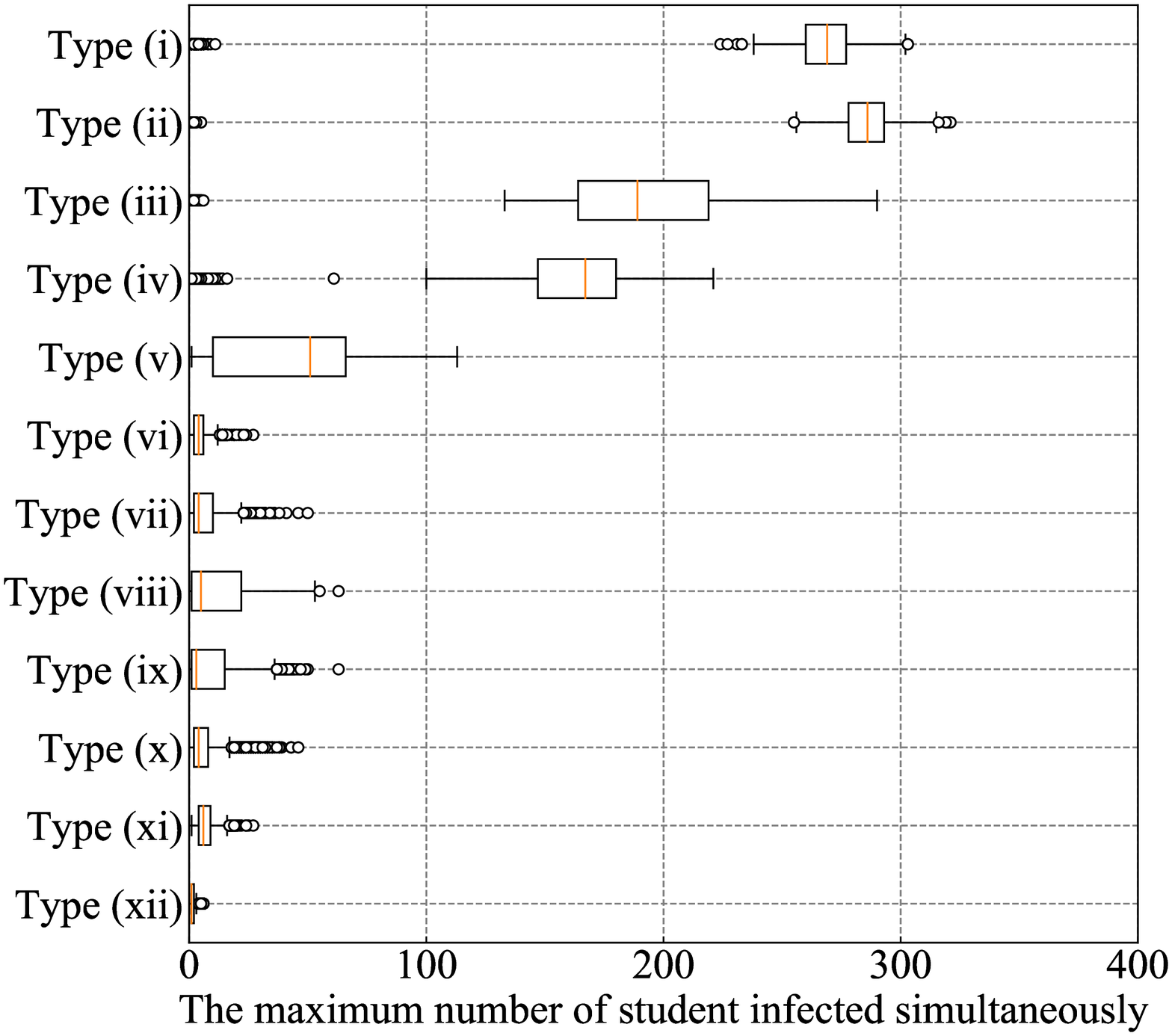}
  \caption{Experiment 2: The maximum number of students infected simultaneously in departmentalized classrooms}
  \label{fig:TheMaxNumHighSchoolExp2}
\end{figure}

\section{Experiment 2}
We simulated school schedules to evaluate the impact of varied school schedules in self-contained, and departmentalized classrooms.
Fig.~\ref{fig:TheMaxNumMiddleSchoolExp2} and ~\ref{fig:TheMaxNumHighSchoolExp2} present the results.
As a general tendency, the maximum number of students infected simultaneously decreases as the percentage of face-to-face lessons increases.

\subsection{Effect of Shortened School Schedule}
The maximum number of type (ii) is slightly higher than that of type (i) in self-contained and departmentalized classrooms.
The total lesson time per day is shorter, and the infection risk per day decreases.
However, the continuous period of face-to-face lessons was one day longer.
An additional day would increase the maximum number of infected students infected simultaneously.

\subsection{Effect of 75\% Face-to-face Lesson School Schedules}
The results of the effect of 75\% face-to-face lessons in the school schedules (type (iii) and type (iv)) in self-contained and departmentalized classrooms do not correspond. 
All students went to school for three weeks, after which they took one week off in type (iii).
The SARS-CoV-2 would spread like type (i) during the face-to-face lesson week.

The maximum number of type (iv) was slightly lower than that of types (i) and (iii).
Three of the four groups went to school alternatively, and each group took face-to-face lessons for three weeks in type (iv).
The period of these was the same as that of type (iii), and the number of students who took face-to-face lessons simultaneously was lower than that of types (i) and (iii).
Thus, the reduction in class size would decrease the maximum number of students infected simultaneously of students infected simultaneously.

\subsection{Effect of 50\% Face-to-face Lesson School Schedules}
The 50\% face-to-face school schedules (types (v), (vi), (vii), (viii), (ix), and (x)) are effective in self-contained and departmentalized classrooms.
The decrease in type (v) was smaller than that in types (vi), (vii), (viii),  (ix), and (x).
All the students go to school for one week, after which they take one week off in type (v).
In contrast, half of the students attended school simultaneously in types (vi), (vii), (viii), (ix), and (x).
Thus, the maximum number of students infected simultaneously correlates with the reduction in class size.

\subsection{Effect of 25\% Face-to-face Lesson School Schedules}
The type (xii) school schedule is effective in self-contained and departmentalized classrooms.
One of the four groups went to school for one week, and took leave for three weeks.
When students are infected during the face-to-face lesson week, symptoms appear while staying at home, and they skip the next face-to-face lesson week.
This reduces the maximum number of infected individuals.

The effects of type (xi) were opposite between self-contained and departmentalized classrooms.
All students went to school for one week, after which they took leave for three weeks. Type (xi) school schedule was effective in departmentalized classrooms; however, it was not as effective in self-contained classrooms.
The maximum number from students infected simultaneously of type (xi) in self-contained classrooms is nearly the same as that from type (v), and higher than that of the other 50\% face-to-face school schedules.

Tables~\ref{tab:P25_W_1F-3_middle_School} and ~\ref{tab:P25_W_1F-3_high_School} show the infection probability of type (xi) in each classroom in self-contained and departmentalized classrooms.
The infection probability increased on Day 4, and that of self-contained classrooms was higher than that of departmentalized classrooms.
Student one comes to school from Day 0 (Monday) to Day 1 (Tuesday), and infects other students.
If the student is asymptomatic, the student continues to attend school. The newly infected students become exposed.
They become infectious or asymptomatic two days later, and start to infect other students. The number then becomes the highest on Day 4 (Friday).
All new infectious and asymptomatic students come to the same classroom in self-contained classrooms; meanwhile, the new infectious and asymptomatic students are scattered across several classrooms in departmentalized classrooms.
Then, the infection probability of classrooms in self-contained classrooms is higher than that in departmentalized classrooms.

\begin{table}[t]
  \caption{Infection probability of type (xi) in each classroom in self-contained classrooms}
  \label{tab:P25_W_1F-3_middle_School}
\begin{tabular}{crrrrrrr}\toprule
                                                                         & \multicolumn{7}{c}{Day}                                                                                                                                                                                                     \\
\begin{tabular}[c]{@{}c@{}}Infection\\ Probability\end{tabular} & \multicolumn{1}{c}{0} & \multicolumn{1}{c}{1} & \multicolumn{1}{c}{2} & \multicolumn{1}{c}{3} & \multicolumn{1}{c}{4} & \multicolumn{1}{c}{5} & \multicolumn{1}{c}{6} \\ \midrule  \midrule
0                                                               & 0\%        & 0\%        & 70\%       & 70\%       & 10\%       & 100\%      & 100\%      \\
0-2\%                                                           & 100\%      & 100\%      & 30\%       & 30\%       & 24\%       & 0\%        & 0\%        \\
2-4\%                                                         & 0\%        & 0\%        & 0\%        & 0\%        & 51\%       & 0\%        & 0\%        \\
4-6\%                                                         & 0\%        & 0\%        & 0\%        & 0\%        & 14\%        & 0\%        & 0\%        \\
6-8\%                                                         & 0\%        & 0\%        & 0\%        & 0\%        & 1\%        & 0\%        & 0\%        \\
8-10\%                                                        & 0\%        & 0\%        & 0\%        & 0\%        & 0\%        & 0\%        & 0\%        \\
10-12\%                                                       & 0\%        & 0\%        & 0\%        & 0\%        & 0\%        & 0\%        & 0\%       \\ \bottomrule                 
\end{tabular}
\end{table}

\begin{table}[t]
  \caption{Infection probability of type (xi) in each classroom in departmentalized classrooms}
  \label{tab:P25_W_1F-3_high_School}
\begin{tabular}{crrrrrrr}\toprule
                                                                         & \multicolumn{7}{c}{Day}                                                                                                                                                                                                     \\
\begin{tabular}[c]{@{}c@{}}Infection\\ Probability\end{tabular} & \multicolumn{1}{c}{0} & \multicolumn{1}{c}{1} & \multicolumn{1}{c}{2} & \multicolumn{1}{c}{3} & \multicolumn{1}{c}{4} & \multicolumn{1}{c}{5} & \multicolumn{1}{c}{6} \\ \midrule  \midrule
0                                                               & 95\%        & 95\%        & 99\%       & 99\%       & 90\%       & 100\%      & 100\%      \\
0-2\%                                                           & 5\%      & 5\%      & 1\%       & 1\%       & 10\%       & 0\%        & 0\%        \\
2-4\%                                                         & 0\%        & 0\%        & 0\%        & 0\%        & 1\%       & 0\%        & 0\%        \\
4-6\%                                                         & 0\%        & 0\%        & 0\%        & 0\%        & 0\%        & 0\%        & 0\%        \\
6-8\%                                                         & 0\%        & 0\%        & 0\%        & 0\%        & 1\%        & 0\%        & 0\%        \\
8-10\%                                                        & 0\%        & 0\%        & 0\%        & 0\%        & 0\%        & 0\%        & 0\%        \\
10-12\%                                                       & 0\%        & 0\%        & 0\%        & 0\%        & 0\%        & 0\%        & 0\%       \\ \bottomrule                  
\end{tabular}
\end{table}

\section{Discussion}
Experiment 1 shows that increasing classroom ventilation rate is effective, as also recommended by the CDC~\cite{CDC21Transmission}.
However, it is found that the impact is not stable when the classroom ventilation rate increases as opposed to customizing school schedules (Experiment 2).

Experiment 2 shows that school schedules can differently impact the maximum number of students infected simultaneously, depending on whether classrooms are self-contained or departmentalized.

UNESCO suggests four types of school schedules - Options 1, 2A, 2B, and 3 - corresponding to school scheduling types (ix), (viii), (vii), and (vi). UNESCO states that infection risk decreases in this order~\cite{UNESCO20}.
UNESCO’s school schedules actually decrease the maximum number of students infected simultaneously, both in self-contained and departmentalized classrooms.
However, we found no significant difference between their effects.

The percentage of class members that the students meet in face-to-face lessons of type (x) is 100\%.
That of UNESCO’s school schedules is 50\%.
In addition, type (x) of both self-contained and departmentalized classrooms has the same effect, as compared to UNESCO’s school schedules.
This means that type (x) of departmentalized classrooms has two advantages - a lower maximum number of students infected simultaneously, and constant interaction with a wide variety of peers.

It was found that type (xi) in self-contained classrooms had a slightly higher maximum number of students infected simultaneously, compared to schedules with a higher percentage of face-to-face lessons.
This result is caused by a combination of the school schedule, the exposed days, and the infectious-exposed days.

These results imply that teachers and education policymakers have to consider a combination of school schedules, and their classroom types; not just the percentage of face-to-face lessons. This is a complex phenomenon, and a difficult task.

\section{Conclusion}
We developed the SVIS for teachers and education policymakers.
It simulates the spread of infection at a school, considering the students’ lesson schedules, classroom volume, air circulation rates in classrooms, and infectability.
We then show the effects of changing classroom volumes, and classroom air change rates, and demonstrate the impact of several school schedules in self-contained and departmentalized classrooms.

The results show that internal school infection is very complex, and can’t obtain the expected result without appropriate school scheduling.
Although, teachers can acquire information like the ranking of four types of classroom models as per their infection risk ~\cite{UNESCO20}, or that ``self-contained classrooms reduce student interaction''~\cite{CTC21Educator}, the SVIS and the simulation results can help teachers and education policymakers plan school schedules appropriately, and reduce the maximum number of simultaneously infected students, while also conducting face- to-face lessons.

However, this study has certain limitations.
There are three main transmission routes of COVID-19~\cite{CDC21Transmission} - inhalation, deposition, and touching.
We focus on inhalation because the airborne (inhalation and deposition) route is estimated as the dominant route for SARSCoV-2 transmission~\cite{zhang2021evidence}.
In addition, deposition has a negligible effect compared with the removal effect of ventilation ~\cite{Dai20}; Wells-Riley equation can consider deposition by treating the ventilation rate (Q) (\ref{eq:p}) as the equivalent clean air delivery rate, which is equal to the actual ventilation rate multiplied by filtration efficiency~\cite{Foarde99}.
It is beneficial to give teachers and school policymakers a reliable, convenient, and simple model with demonstrated results.
In addition, the SVIS does not consider effective measures in reducing the transmission of COVID-19, such as frequently conducting antigen and PCR tests, contact tracing, cleaning, and disinfecting~\cite{Ghaffarzadegan21,CDC21Cleaning}.
In the future, the SVIS should consider these factors.



\bibliography{sample}
\bibliographystyle{unsrt}

\end{document}